\documentstyle[twoside,a4,12pt]{article}
\input epsf
\epsfverbosetrue
\def\simge{\rlap{\raise 2pt \hbox{$>$}}{\lower 2pt \hbox{$\sim$}}}
\def\simle{\rlap{\raise 2pt \hbox{$<$}}{\lower 2pt \hbox{$\sim$}}}
\begin{document}
%
%
\begin{center}
{\footnotesize
To be published in the Proceedings of the International Seminar on
Current Developments in Disordered Materials, Kurukshetra University,
India, 22--24 January 1996.\\[0.5cm]
}
{\Large \bf Diffusion Coefficients of Single and Many Particles in Lattices
with
Different Forms of Disorder}\\[0.3cm]

K.W. Kehr and T. Wichmann\\[0.3cm]
Institut f\"ur Festk\"orperforschung, Forschungszentrum J\"ulich GmbH, 52425
J\"ulich, Germany\\
\vspace{0.5cm}
{\sl Topics: random transition rates, random barriers, random traps,
Miller-Abrahams model,
 diffusion coefficient,
coefficient of collective diffusion, effective medium theory,\\
 site-exclusion model}\\
\end{center}

{\bf Abstract}

A survey is given on asymptotic diffusion coefficients of particles in lattices
with random transition rates.
Exact and approximate results for single particles are reviewed. A recent exact
expression in $d = 1$ which includes
occupation factors is discussed. The utilization of the result is demonstrated
for the Miller-Abrahams model and a model
of random barriers combined with random traps. Exact and approximate results
for the site-exclusion model in disordered
lattices are also given.\\

{\bf 1. Introduction}

Transport processes of particles in disordered materials exhibit a variety of
interesting phenomena, such as a strong
reduction of the asymptotic diffusion coefficients, anomalous frequency
dependence of the conductivity, dispersive
transport, etc. The explanation of the transport processes of single and many
particles in disordered materials has been a challenge
to theory; it is of great practical importance as well. This paper focuses on
the asymptotic diffusion
coefficients of single and many particles for different models of disordered
lattices. Quite different models of disorder
were introduced to describe particle transport in crystals with point defects,
as well as in amorphous materials and
glasses. Exact results will be presented as far as possible; these results are
mainly but not always restricted to the dimension
$d = 1$. Recently a general exact expression for the diffusion coefficient of
single particles in $d = 1$ has been derived \cite{wich,diet,kut}.
The insight obtained from this derivation can serve as a basis for the
effective-medium approximation in higher dimensions.\par
The derivation of diffusion coefficients of many particles where multiple
occupancy of sites is excluded will be restricted
to the coefficient of collective diffusion which appears in Fick's law. Exact
results for this diffusion coefficient in disordered
lattices are scarce. It will be shown that a properly formulated
effective-medium approximation yields reasonable results
in higher-dimensional disordered lattices. \\

{\bf 2. Models}

Most theoretical models for the description of transport in disordered
materials use ordered lattices and put the disorder
into the transition rates between the sites. Typically only nearest-neighbor
transitions are considered. Several models of the
disorder have been considered in the past. Two prototype models are the
random-barrier (RB) model and the random-trap (RT) model \cite{HK},
see Fig.\ 1 (a) and (b). In the RB model the transition rates between two
neighbor sites $i$ and $j$ are symmetric, $\Gamma_{ij} = \Gamma_{ji}$.
In the RT model, the transition rates $\Gamma_i$ originating from the
sites are given; they are independent of the final sites.\par
The transition rates $\Gamma_{ij}$, or $\Gamma_i$, respectively, are taken from
a common distribution, $f(\Gamma)$. Usually thermal
activation is assumed as the physical process leading to transitions. The
simplest form of a thermally activated process is given by the
Arrhenius law, $\Gamma = \Gamma_0 Ê\exp (-E/k_BT)$.
\pagebreak[4]
%
%
\textheight25.cm
In the case of the RB model, the energy $E$ represents the barrier height,
while for the
RT model
$E$ is indicative of the trap depth. Instead of specificing the distribution
$f(\Gamma)$, one may specify the energy distribution
$\nu(E)$, from which barrier heights for the RB model, or trap depths of the RT
model are chosen. Note that all sites have
the same energy in the RB model, while all barriers lie at the same energy in
the RT model.\par
\begin{figure}[htb]
\epsffile{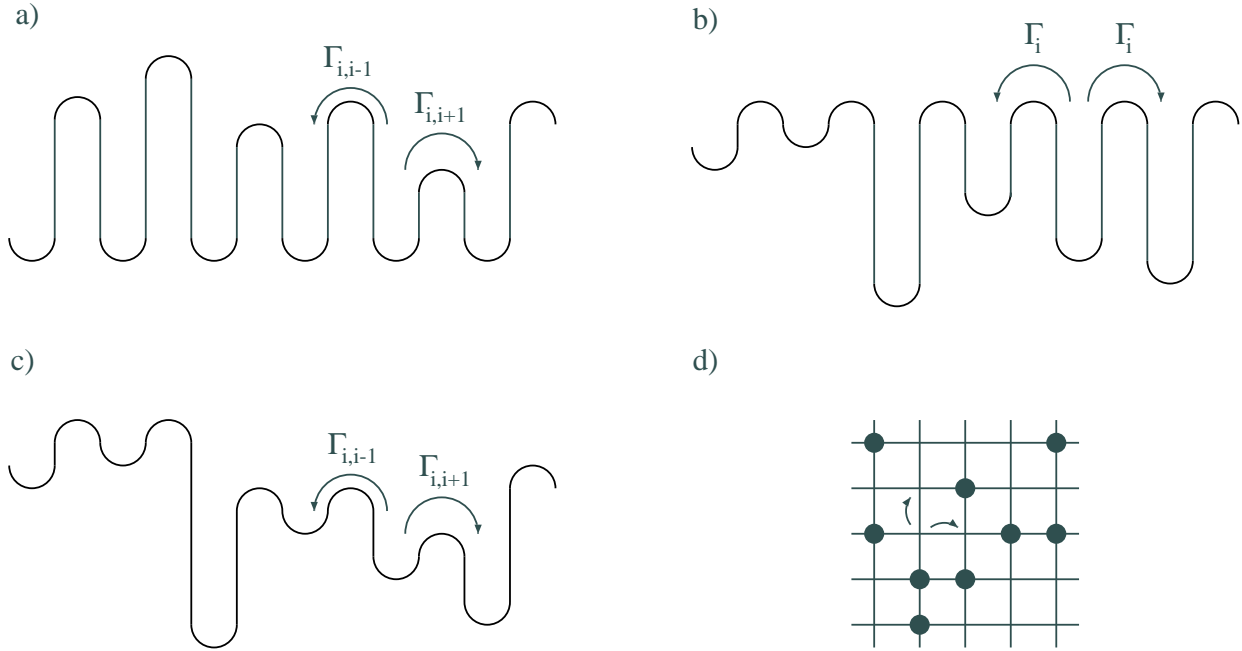}
\caption[]{Schematic representation of the four models of disordered transition
rates that are discussed in the text.}
\end{figure}
A third model is the one introduced by Miller and Abrahams (MA) \cite{MA}, see
Fig.\ 1(c). In the MA model all transitions to neighbor sites have identical
rates, if the final site has a lower energy than the initial site. Transitions
that lead to energetically higher sites require thermal
activation,
\begin{equation}
\Gamma_{i\to j} = \left\{ \begin{array}{lr} \Gamma_0 & E_j \le E_i\\ \Gamma_0
\exp (\frac{E_i-E_j}{k_BT}) & E_j \ge E_i \end{array} \right. . \label{mam}
\end{equation}
The model was introduced for electron or hole transport in amorphous
semiconductors. For this problem a distance-dependent factor
appears in the transition rate. It seems reasonable for particle diffusion to
restrict the modeling to nearest neighbor transitions.\par
For completeness also the model of randomly blocked sites (RBS) is considered
\cite{stauf}, cf. Fig.\ 1(d). Particles are allowed to make transitions between
accessible sites, with a uniform transition rate $\Gamma$. If the concentration
$p$ of open sites becomes smaller than the percolation concentration
$p_c$, long-range diffusion becomes impossible. We restrict the discussion to
concentrations $p > p_c$ where an asymptotic diffusion
coefficient exists.\\
\voffset0.1cm

{\bf 3. Previous Results for Single Particles}

What is known about the asymptotic diffusion coefficients of single particles
in these different models? An exact result is known for the RB model
in $d = 1$, cf. \cite{alex},
\begin{equation}
D = \left\{\frac{1}{\Gamma} \right\}^{-1}.  \label{rbd1}
\end{equation}
The curly brackets in Eq.\ \ref{rbd1} denote the disorder average over the
distribution $f(\Gamma)$ of the transition rates. For
simplicity the lattice constant has been set to unity. A consequence of Eq.\
\ref{rbd1} is that the diffusion coefficient
vanishes when the average diverges. This is the case for an exponential
distribution of barrier heights, $\nu(E) = \sigma^{-1} \exp (-E/\sigma)$
at temperatures where $k_BT \le \sigma$.\par
For the RB model in higher dimensions, one has to resort to approximations. One
popular approach is the effective-medium
approximation (EMA). An effective transition rate is determined from a
self-consistency condition, which reads \cite{kirk,berna}
\begin{equation}
\left\{ \frac{\Gamma_{eff} - \Gamma}{\frac{z-2}{2} \Gamma_{eff} + \Gamma}
\right\} = 0 \label{self}
\end{equation}
where $z$ is the coordination number of the lattice. Again the average is taken
over the distribution of the rates $f(\Gamma)$. The
diffusion coefficient is then identical to $\Gamma_{eff}$. Usually
the EMA gives good results in $d = 3$ for not too strong disorder
\cite{amb}.\par
Another estimate of the asymptotic diffusion coefficient
is provided by the critical-path (CP) approach of Ambegaokar, Halperin, and
Langer \cite{AHL}. In this approach the path from one
to the other side of a large lattice is considered that overcomes the lowest
possible barriers. The highest barrier in this path determines
the asymptotic diffusion coefficient, and it is related to the bond percolation
threshold. The CP approach is appropriate for strong
disorder in $d = 3$ \cite{tyc}. For simple square lattices EMA and CP yield the
same result, for a uniform distribution of activation energies
between 0 and $E_c$,
\begin{equation}
D = \Gamma_0 \exp \left(-\frac{E_c}{2k_BT}\right).
\end{equation}
This is in fact an exact result, as a consequence of the self-duality of this
lattice \cite{berna}.\par
An exact result is known for the RT model {\sl in all dimensions}. It reads
\begin{equation}
D = \left\{ \frac{1}{\Gamma} \right\}^{-1}   \label{rtd}
\end{equation}
where hypercubic lattices have been assumed and the lattice constant is unity.
This result was first derived in  \cite{schr}.
That it is exact, can be shown by establishing the strict linearity of the
mean-square displacement for all (positive) times, for equilibrium
initial conditions \cite{HKL}. The physical reason for this result is the
absence of any time-dependence in the
correlations of consecutive transitions. Equation 5 follows then easily from
the short-time behavior.\par
For the MA model to our knowledge no exact result has been published for $d =
1$. The interest in this model concentrated on the
physically relevant dimension $d = 3$. Here EMA results have been developed,
see the review \cite{BB}.\par
Finally, for the model of randomly blocked sites an approximate result for
small concentrations of blocked sites has been derived
by Tahir-Kheli \cite{take}. It reads
\begin{equation}
D = \Gamma \left(1 - \frac{1-p}{f} + \cdots\right)   \label{rbs}
\end{equation}
where $p$ is the concentration of accessible sites and $f$ the correlation
factor for tracer diffusion in lattice gases, in the limit of
concentration $c \to 1$. Equation \ref{rbs} represents an expansion in $1 - p$
and the next term is known in $d = 2$ \cite{ernst}.
Numerical simulations \cite{braun} demonstrated that Eq.\ \ref{rbs} is a very
good approximation in $d = 3$ for $p \,\simge \, 0.8$.\\

{\bf 4. Exact Result for $d = 1$}

Recently an exact expression for the asymptotic diffusion coefficient in
one-dimensional lattices has been derived \cite{wich,diet,kut},
which is valid for arbitrary forms of the disordered transition rates. The
assumption has to made that an equilibrium state exists for the site
occupancy in the limit of infinite chains. One approach for the derivation of
the expression is to consider the mean first-passage time
(MFPT) of a particle from a starting-site 0 to a terminal site $N$ on a segment
of a disordered chain \cite{wich}. An exact expression is known
for the MFPT $\bar{t}_{0N}$, for fixed disorder, in which all transition rates
on the segment appear explicitly \cite{MK},
\begin{equation}
\label{gt0n}
\bar{t}_{0N} =
\sum_{k=0}^{N-1}\frac{1}{\Gamma_{k, k+1}}+
\sum_{k=0}^{N-2}\frac{1}{\Gamma_{k, k+1}}
\sum_{i=k+1}^{N-1}\prod_{j=k+1}^{i}
\frac{\Gamma_{j, j-1}}{\Gamma_{j, j+1}}.
\end{equation}
If an equilibrium state exists, the relation of detailed balance is valid
between neighbor sites,
\begin{equation}
\label{gdb}
\rho_i\Gamma_{ij} = \rho_j\Gamma_{ji}\,\,\,\,\,\,\, {\rm with} \,\,\,\,\,\,\,
\rho_i = \frac{\exp (-\beta E_i)}{\{\exp (-\beta E_i)\}}
\end{equation}
and $\beta = (k_BT)^{-1}$. Here $\rho_i$ is an occupation factor, which is
proportional to the
occupation probability of site $i$.
Equation \ref{gdb} can be used to simplify the second double sum
in Eq.\ \ref{gt0n}. One obtains as the disorder average of the MFPT
in the long segment limit $N\to\infty$
\begin{equation}
\label{gt0nlimit}
\left\{ \bar{t}_{0N}\right\} =
\frac{1}{2}N^2\left\{\frac{1}{\rho_i \Gamma_{ij}}
\right\}.
\end{equation}
Comparison of Eq.\ \ref{gt0nlimit} with the inverted asymptotic relation
between time and mean-squared displacement $t = (2D)^{-1}\{x^2\}$
yields the final result
\begin{equation}
\label{dex1}
D = \left\{ \frac{1}{\rho_i \Gamma_{ij}} \right\}^{-1}.
\end{equation}

By analogous derivations as sketched above one can show that the mean-squared
deviation of the mean-first passage time behaves as
\begin{equation}
\{ \bar{t}^2_{0N}\} - \{ \bar{t}_{0N}\}^2 \propto N^3.
\end{equation}
This result means that the relative deviations from the mean value become small
for large $N$. In this sense it is justified to use
the inverted connection between time and distance-squared to deduce the
diffusion coefficient.\par
Equation \ref{dex1} has been derived in \cite{diet,kut} from the linear
response of the mean particle current to
an external driving force, invoking the Einstein relation between mobility and
diffusivity.\par
The physical significance of the result Eq.\ \ref{dex1} is that the diffusion
coefficient on disordered linear chains is determined
by transition rates that are weighted by the thermal occupation of the initial
sites. Loosely speaking, the diffusion coefficient
follows from the actual transition rates of the particles.\par
It will be first verified that Eq.\ {\ref{dex1} reproduces previous exact
results. This is trivial for the RB model, where
all $\rho_i = 1$, and Eq.\ \ref{rbd1} is immediately obtained. Since particles
in the RT model require thermal activation that is given
by the energetic depth of the sites, measured from a common origin, the
occupation factors are inversely proportional to the
transition rates, $\rho_i \sim \Gamma^{-1}_i$. Equation \ref{dex1} then reduces
to
\begin{equation}
D =  \Gamma_0 \{ \exp (-\beta E)\}^{-1}.   \label{rtnew}
\end{equation}
In view of the equivalence between $\rho_i$ and $\Gamma_i^{-1}$, Eq.\
\ref{rtnew} is completely equivalent to Eq.\ \ref{rtd}. The
expression Eq.\ \ref{rtnew} provides a new {\sl interpretation} of the general
result Eq.\ \ref{rtd} for the RT model. In this model,
the diffusion coefficient is reduced in accordance with the distribution of the
occupation factors. If these are widely distributed,
the diffusivity is reduced accordingly.\par
A nontrivial application of Eq.\ \ref{dex1} is provided by the MA model, in the
form presented in Eq.\ \ref{mam}. The resulting
diffusion coefficient is
\begin{equation}
D = \frac{1}{2} \Gamma_0 \{ \exp (-\beta E_i)\}^{-1} \{ \exp (\beta E_f)
\}^{-1}_{E_f \ge E_i}.
\end{equation}
The second average is actually a double average over distributions $\nu(E_i)$
and $\nu(E_f)$. Fig.\ 2 gives a numerical
verification for the discrete two-level distribution of transition rates.\\

\begin{figure}[htb]
\epsffile{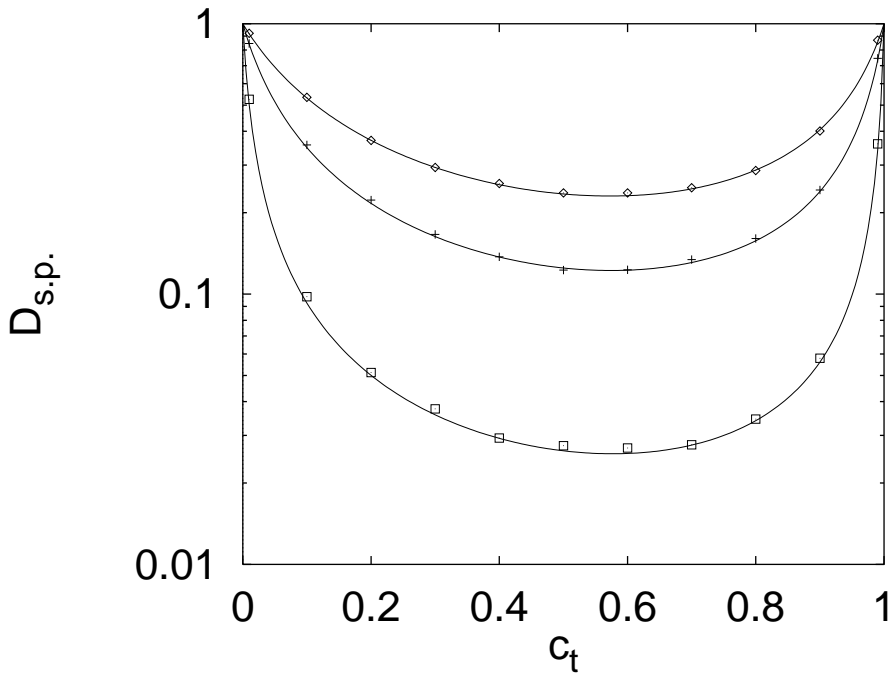}
\caption[]{Diffusion coefficient of single particles in a two-level MA model in
$d = 1$ with $c_t$ the concentration of lower sites.
The up transition rates are 0.1 ($\Diamond$), 0.05 (+), and 0.01 ($\Box$) in
units of $\Gamma_0$.}
\end{figure}

{\bf 5. Effective-Medium Approximation for $d > 1$}

To derive diffusion coefficients of particles in higher-dimensional disordered
lattices one has to resort to approximations.
Equation \ref{dex1} gives a suggestion how to handle different forms of
disorder, when thermal occupancy of the sites plays a
role. The suggestion is to use transition rates that are weighted by the
thermal occupation factors. In the further derivations
the following rates will be used
\begin{equation}
\Gamma^{sym}_{ij} = \rho_i\Gamma_{ij} .   \label{sym1}
\end{equation}
The transition rates Eq.\ \ref{sym1} are labeled "sym" because
they are symmetric in view of the relation of detailed balance. This is of
direct relevance, since the self-consistency condition
Eq.\ \ref{self} requires symmetric transition rates. The use of Eq.\ \ref{sym1}
in the self-consistency condition Eq.\ \ref{self} constitutes
the EMA for the different models of disorder.\par
The RB model is trivial in the sense that all $\rho_i = 1$ and the
self-consistency condition can be evaluated with the original
symmetric rates. Various results for the RB model have been published in the
past, see, for instance \cite{arg}. Evaluation
of the self-consistency condition Eq.\ \ref{self} for the RT model reproduces
the dimension-independent exact result Eq.\ \ref{rtd}. A nontrivial
example is provided by the MA model, where results for a Gaussian distribution
of site energies in $d = 3$ are presented in Fig.\ 3.
Deviations between the EMA and the numerical simulations are visible for larger
disorder.\\

\begin{figure}[htb]
\epsffile{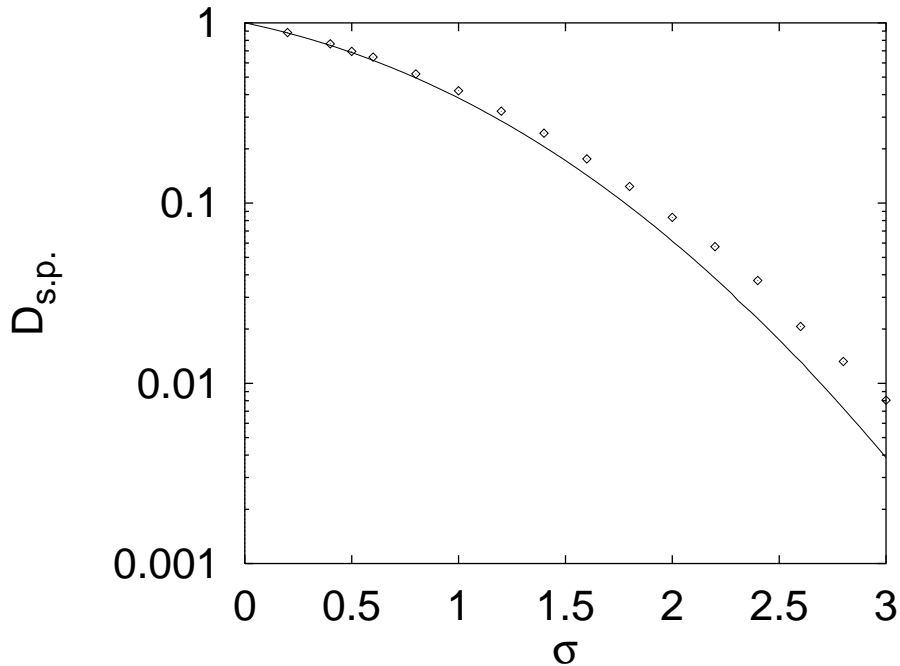}
\caption[]{Diffusion coefficient of single particles in the MA model in $d = 3$
for Gaussian distributions of the energies.
The width $\sigma$ of the Gaussian is given in units of $k_BT$.}
\end{figure}

{\bf 6. Combination of random barriers and random site energies}

In various amorphous materials the observed diffusion coefficients
exhibit Arrhenius behavior, i. e. $\ln D$ vs $1/T$ shows linear
behavior. This is surprising because the different models discussed above yield
curvature in an Arrhenius plot of $D$.
The RT model leads generally to downward curvature, an example will be given
below. The RB model leads to upward curvature
in $d = 3$, an example is given in \cite{arg}. Limoge and Bocquet \cite{LB}
suggested that in a combination of the RB and the RT model
cancelation of both effects may occur, resulting in an Arrhenius behavior of
the diffusion coefficient. The methods
outlined above are well suited to treat this problem.\par
Equation \ref{dex1} shows that the combination $\rho_i\Gamma_{ij}$ enters into
the theory. The energy to be surmounted
from a site over a barrier leading to the neighbor site is the sum of a site
energy and a barrier energy. The two random
quantities are assumed to be independent. In the combination $\rho\Gamma$, the
contributions of the site energies cancel
each other, as for the pure RT model, while the barrier energy remains
unaffected. As a result, the diffusion coefficient
of the combined model is
\begin{equation}
D_{comb} = \frac{1}{\Gamma_0}D_{RT} D_{RB}.
\end{equation}
The coefficient $D_{RT}$ is given by Eq.\ \ref{rtnew} in any dimension.
$D_{RB}$ can be evaluated exactly in $d = 1$, cf. Eq.\ \ref{rbd1},
and approximately from the EMA in higher dimensions.\par
Since the expressions Eqs. \ref{rbd1} and \ref{rtd} are identical in $d = 1$,
the effect of random barriers is the same as that of
random site energies, if they have the same distribution. As a result, no
compensation effect will occur in $d = 1$, and the
downward curvature of $D$ in the Arrhenius plot of the RT model will be
enhanced by random barriers. Compensatory effects are
possible in $d = 3$, if the strength of the disorder is adjusted accordingly.\\

\begin{figure}[htb]
\epsffile{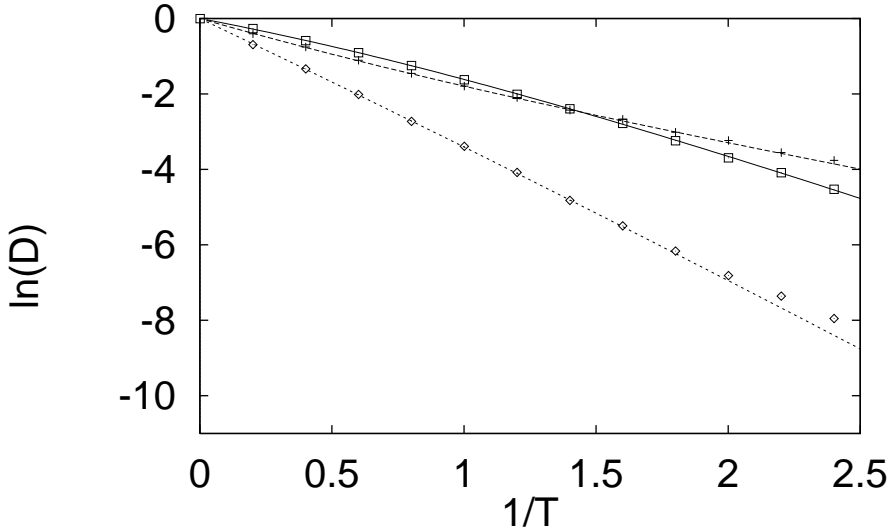}
\caption[]{Diffusion coefficient of single particles in the combined RT and RB
model, for uniform distributions of the energies.
Numerical results for the RB (+), RT ($\Box$), and the combined model
($\Diamond$) are shown together with the EMA results.}
\end{figure}
The conclusion can be drawn that cancelation of the effects of random site
energies and of random barriers is possible, under
the assumption of independent distributions of the two quantities. However, the
cancelation requires a specific choice of the parameters
of the distributions. More details for this model will be published elsewhere
\cite{mussa}.\\

{\bf 7. Many Particles: Previous Exact Results}

Now many particles in lattices with disordered transition rates will be
considered. Multiple occupancy of the sites will be
forbidden, otherwise the particles do not have interactions. The derivations
will be restricted to the coefficient of
collective diffusion, which appears in Fick's law. That is, the decay of
density deviations is considered, also in the simulations,
on a mesoscopic scale. If this decay is exponential for longer times, with a
coefficient inversely proportional to the wavelength
squared, the coefficient of collective diffusion can be deduced. \par
There are only a few exact results available for the collective diffusion
coefficient of lattice gases on disordered lattices. The most
prominent result is that the coefficient of collective diffusion agrees with
the diffusion coefficient of single particles whenever all transition rates
between two neighbor sites are symmetric. There is a cancelation of the
exclusion effects of
double occupancy, leading to $D_{coll} = D_{s.p.}$ \cite{kut1}. This result
holds for the RB model and the model of randomly blocked sites. A
numerical verification of it was made for the RBS model in \cite{braun}. A
consequence of the result is that the derivation
of $D_{coll}$ becomes nontrivial for the
RT and MA models, and for combinations involving the RT model. A trivial
limiting case is the one of low concentrations of particles,
in which $D_{coll} \to D_{s.p.}$.\par
Lattice gases in the RT model can be considered, in the limit of concentration
$c \to 1$, as models with a low concentration of vacancies in
random lattices, whose transition rates {\sl into} the sites are specified. The
diffusion coefficient can be calculated exactly in this
limit in disordered linear chains \cite{paetz} and the result is
\begin{equation}
D_{coll} = \left\{ \frac{1}{\Gamma} \right\}^{-2} \left\{\Gamma\right\}^{-1}.
\label{dcv1}
\end{equation}
The result Eq.\ \ref{dcv1} implies that in a two-level model with \lq\lq
free\rq\rq \, and \lq\lq trap\rq\rq \, sites, the diffusion coefficient has a
minimum for
a trap concentration of $c_t = 2/3$, not at the concentration $c_t = 1$ of
sites of low energies.\par
Another limit that is well understood
is the two-level random trap model for $d > 1$ when the concentration of trap
sites is small and the free sites are percolating.
If the hopping rate out of the trap sites is made very small, and if $c > c_t$,
the trap sites are saturated by particles
that are practically immobile. This is effectively the RBS model, where the
collective diffusion coefficient is given by the single-particle
diffusion coefficient. It was pointed out in Sec.\ 3 that Eq.\ \ref{rbs} is a
very good approximation for the single particle diffusion
coefficient in this model. Numerical simulations \cite{kehr,born} verified that
the coefficient of collective diffusion in the situation
described above is given by Eq.\ \ref{rbs}.\\

{\bf 8. Effective-Medium Approximation at Finite Particle Concentrations}

Apart from the special situations described above, approximate methods are
necessary to derive the collective diffusion coefficient
for the RT and MA models. For arbitrary dimensions one would like to extent the
EMA to finite particle concentrations. This requires
in a first step the introduction of effective or mean-field single-particle
transition rates. Single-particle transition rates can be
introduced in two different forms. One possibility is to factorize the
two-particle probabilities that occur in the master equation and to
linearize around the equilibrium solution \cite{paetz}. The derivation of Eq.\
\ref{dcv1} was based on this procedure. The second
possibility is to introduce symmetrized transition rates by \cite {piga}
\begin{equation}
\Gamma^{Sym}_{ij} = \frac{P_i(1-P_j) \Gamma_{ij}}{\{P_i(1-P_i)\}}.
\label{sym2}
\end{equation}
Here $P_i$ are the site occupation probabilities in equilibrium, with $0 < P_i
< 1$. The rates Eq.\ \ref{sym2} are symmetric as a consequence
of the detailed balance condition for lattice gases. They are generalizations
of the rates Eq.\ \ref{sym1} for single particles; note that the
$P_i$ and $\rho_i$ are differently normalized. The denonimator in Eq.\
\ref{sym2} has been introduced for normalization.\par
The result for $D_{coll}$ in $d = 1$ that follows from Eq.\ \ref{sym2} will be
discussed in the next section. In higher dimensions, the
symmetrized rates can be used in the self-consistency condition Eq.\
\ref{self}; this constitutes an EMA for lattice gases
on disordered lattices. Numerical simulations for the RT model were compared
with the EMA in Ref. \cite{born,wich2}. There is good
agreement between the EMA and the simulations for smaller $c$, and rough
qualitative agreement for larger $c$. Also the single-particle
limit is obtained correctly.\par
Results for the MA model are presented in Fig.\ 5. The exponential distribution
was taken for the site energies, and the relevant parameter
is the ratio between the width $\sigma$ and the temperature. This distribution
is interesting because the single-particle diffusion
coefficient vanishes also in the MA model for $k_BT \le \sigma$. Due to the
saturation of the low-energy states a coefficient of collective
diffusion exists for finite concentrations. For $k_BT < \sigma$ and small $c$ a
power-law dependence holds,
\begin{equation}
D \approx \Gamma_0 \left(\frac{\sigma}{k_BT} - 1\right) c^{(\sigma/k_BT-1)},
\end{equation}
as has been found for the RT model \cite{wang}. The simulations agree with the
EMA for small $c$, there are deviations at larger $c$.\\

\begin{figure}[htb]
\epsffile{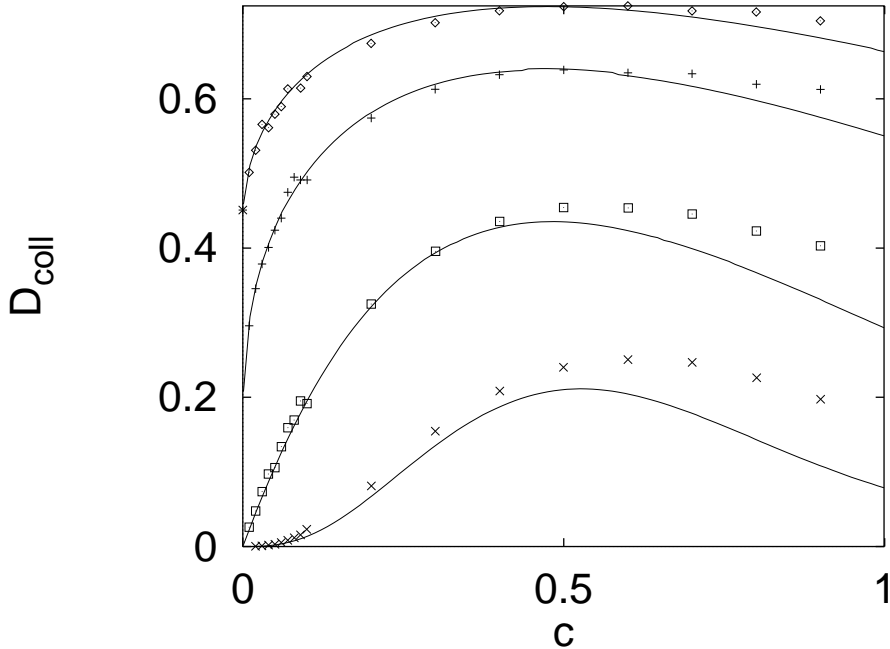}
\caption[]{Collective diffusion coefficient in the MA model in $d = 3$, with an
exponential energy distribution. The ratio $\sigma/k_BT$ was
0.7 ($\Diamond$), 1 (+), 2 ($\Box$), and 4 ($\times$). The symbols are
simulation results and the lines represent the EMA. $D_{s.p.}$ for
$\sigma/k_BT$ = 0.7
is indicated on the ordinate.}
\end{figure}
In the limit of infinite coordination number $z \to \infty$, the EMA yields
\begin{equation}
D^{phen}_{coll} = \left\{Ê\Gamma^{Sym}_{ij} \right\}.
\end{equation}
This expression was previously derived as a phenomenological theory of
lattice-gas diffusion on disordered lattices
\cite{gries,sal}. Further details are given in \cite{wich2}.\\

{\bf 9.  Open Problems}

Missing parts of a complete theory of diffusion of many particles in disordered
lattices are evident. Interaction of particles has been neglected
here, except the exclusion of double occupancy of sites. Also tracer diffusion
has been disregarded. Tagged-particle motion involves
additional correlations, which are certainly difficult to treat in disordered
lattices. Disorder has been restricted to local defects
(sites or bonds); the inclusion of extended defects would be of great interest.
\par
How far is collective diffusion of site-exclusion lattice gases in disordered
lattices understood? One can say that it is understood from a practical point
of view. The major qualitative effect is the saturation of trapping sites.
Approximate methods exist to deal with the problem in higher
dimensions. From a fundamental point of view, however, the problem is not yet
solved. This becomes apparent in $d = 1$. From the symmetrized rates
Eq.\ \ref{sym2} the following expression is obtained in $d = 1$ \cite{piga}:
\begin{equation}
D_{coll} = \left\{ (\Gamma_{Sym})^{-1} \right\}^{-1}.  \label{cd1}
\end{equation}
An equivalent result follows from the
unsymmetrized mean-field rates \cite{paetz}.The result shows strong
discrepancies with the numerical simulations at intermediate particle
concentrations.
The discrepancies can be traced back to the neglect of two-particle
correlations when the
single-particle rates are introduced. Garter and Pitis \cite{piga} calculated
corrections to Eq.\ \ref{cd1}; there remain still discrepancies. The
conclusion is that the two-particle correlations are not understood in
time-dependent situations. There remains the task to develop
a correct theory for this problem, at least in $d = 1$.


\newpage
\begin{thebibliography}{xxx}
\parsep0.0cm
\itemsep0.0cm
\bibitem{wich}
T. Wichmann, PhD thesis, in preparation.
\bibitem{diet}
W. Dieterich, unpublished.
\bibitem{kut}
R. Kutner, Physica A, in the press.
\bibitem{HK}
J.W. Haus and K.W. Kehr, Phys. Rep. {\bf 150}, 263 (1987).
\bibitem{MA}
A Miller and E. Abrahams, Phys. Rev. {\bf 120}, 745 (1960).
\bibitem{stauf}
D. Stauffer, {\sl Introduction to Percolation Theory}, Taylor \& Francis,
London (1985).
\bibitem{alex}
S. Alexander, J. Bernasconi, W.R. Schneider, and R. Orbach, Rev. Mod. Phys.
{\bf 53}, 175 (1981).
\bibitem{kirk}
S. Kirkpatrick, Rev. Mod. Phys. {\bf 45}, 574 (1973).
\bibitem{berna}
J. Bernasconi, Phys. Rev. B {\bf 7}, 2252 (1973).
\bibitem{amb}
H. Ambaye and K.W. Kehr, Phys. Rev. E {\bf 51}, 5101 (1995).
\bibitem{AHL}
V. Ambegaokar, B.I. Halperin, and J. Langer, Phys. Rev. B {\bf 4}, 2612 (1971).
\bibitem{tyc}
S.Tyc and B.I. Halperin, Phys. Rev. B {\bf 39}, 877 (1989).
\bibitem{schr}
K. Schroeder, Proc. Int. Conf. on Fundamental Aspects of Radiation Damage in
Metals, ERDA
CONF751006, Vol. 1, eds. M.T. Robinson and F.W. Young Jr. (1975) p. 525.
\bibitem{HKL}
J.W. Haus, K.W. Kehr, and J.W. Lyklema, Phys. Rev. B {\bf 25}, 2905 (1982).
\bibitem{BB}
H. Boettger and V.V. Bryksin, Phys. Stat. Sol. (b) {\bf 113}, 9 (1982).
\bibitem{take}
R.A. Tahir-Kheli, Phys. Rev. B {\bf 28}, 3049 (1983).
\bibitem{ernst}
M.H. Ernst, T.M. Nieuwenhuizen, and P.F.J. van Velthoven, J. Phys. A {\bf 20},
5335 (1987).
\bibitem{braun}
M. Braun and K.W. Kehr, Phil. Mag. A {\bf 61}, 855 (1990).
\bibitem{MK}
K.P.N. Murthy, and K.W. Kehr, Phys. Rev. A {\bf 40}, 2082 (1989); Erratum {\sl
ibid.}{\bf 41}, 1160 (1990).
\bibitem{arg}
P. Argyrakis, A. Milchev, V. Pereyra, and K.W. Kehr, Phys. Rev. E {\bf 52},
3623 (1995).
\bibitem{LB}
Y. Limoge and J.L. Bocquet, Phys. Rev. Lett. {\bf 65}, 60 (1990).
\bibitem{mussa}
K. Mussawisade, T. Wichmann, and K.W. Kehr, in preparation.
\bibitem{kut1}
R. Kutner, Phys. Lett. A {\bf 81}, 239 (1981).
\bibitem{paetz}
K.W. Kehr, O. Paetzold, and T. Wichmann, Phys. Lett. A {\bf 182}, 135 (1993).
\bibitem{kehr}
K.W. Kehr and O. Paetzold, Physica A {\bf 190}, 1 (1992).
\bibitem{born}
K.W. Kehr and T. Wichmann, in {\sl Diffusion Processes: Experiment, Theory,
Simulations},
ed. A. P\c{e}kalski, Lecture Notes in Physics Vol. 438 (Springer, 1994) p. 179.
\bibitem{piga}
P. Gartner and R. Pitis, Phys. Rev. B {\bf 45}, 7739 (1992).
\bibitem{wich2}
T. Wichmann and K.W. Kehr, J. Phys.: Condens. Matter {\bf 7}, 717 (1995).
\bibitem{wang}
T. Wichmann, K.G. Wang, and K.W. Kehr, J. Phys. A {\bf 27}, L263 (1994).
\bibitem{gries}
R.C. Brouwer, E. Salomons, and R. Griessen, Phys. Rev. B {\bf 38}, 10217
(1988).
\bibitem{sal}
E. Salomons, J. Phys. C: Solid State Phys. {\bf 21}, 5953 (1988).
\end{thebibliography}
\end{document}